\documentstyle[epsf,sprocl]{article}

\def\be{\begin{equation}}
\def\ee{\end{equation}}
\def\bea{\begin{eqnarray}}
\def\eea{\end{eqnarray}}
\begin{document}

\title{
Electromagnetic Processes in Three-Nucleon Systems
}

\author{W. Gl\"ockle, H.~Kamada\footnote{present address: Institut f\"ur
    Kernphysik, Fachbereich 5 der Technischen Hochschule Darmstadt, D-64289
    Darmstadt, Germany   }}
\address{ Institut f\"ur Theoretische Physik II,
         Ruhr Universit\"at Bochum, 44780 Bochum, Germany}
\author{J. Golak, H.~Wita\l a}
\address{ Institute of Physics, Jagellonian University,
                    PL 30059 Cracow, Poland}
\author{S. Ishikawa}
\address{ Department of Physics,
       Hosei University,
       Fujimi 2-17-1, Chiyoda,
       Tokyo 102,
       Japan}
\author{D. H\"uber}
\address{ Los Alamos National Laboratory,
Theoretical Division,
M.S. B283,
Los Alamos, NM 87545
USA  }


\maketitle\abstracts{
Results gained for electron scattering on $^3$He in a nonrelativistic
framework are reviewed.
The electromagnetic current is truncated to a single nucleon operator,
but the interaction among the three nucleons is treated exactly.
Thus initial and final 3N states are evaluated consistently as
solutions of Faddeev equations based on realistic NN forces.
The correct inclusion of the final state interaction
turned out to be important. The agreement to data is reasonably good,
but the neglection of MEC's can be seen. That inconsistency
should be removed. For pd capture processes we included some
MEC's via the Siegert theorem, which dramatically improves the
description of the pd capture data.
}

\section{Introduction}
\label{secIN}

Elastic and inelastic electron scattering on $^3$He($^3$H) as well as 
photodisintegration of $^3$He or pd capture have been studied since many 
years \cite{REF1}. One hopes to get insights into the 3N bound state wavefunctions
and into the hadronic current 
operator. The single nucleon momentum distribution, NN correlation functions and d-state admixtures are prominent examples of 3N bound state properties;
electromagnetic form factors of the hadrons (especially the ones of the neutron), the role and properties of two-nucleon currents or even three-nucleon currents and their consistency to the underlying nuclear forces are important issues related to the current operator. 
Last not least for higher energy and momentum transfers brought in by the photon into the nuclear system relativistic effects can no longer be neglected and call for strong efforts to widen the familiar nonrelativistic Schr\"odinger equation into the realm of relativity.
Only in relation to a relativistic formulation of an effective hadronic theory will it be possible to control the transition region from hadronic to quark degrees of freedom. 

The contribution of this article will cover only a small section of this wide arena: the nonrelativistic regime and a still truncated current operator. 
But the results are promising and will serve as benchmarks. 
A very important ingredient is the exact treatment of the interaction among the three nucleons in continuum states, whether it is the final state interaction in electron or photon induced breakup processes of $^3$He or the initial state interaction in pd capture processes. 
This is described in section II in two examples and we refer to \cite{REF2} for more detailed presentations. 
Needless to say that the three-nucleon wavefunctions should be based on realistic nuclear forces.
We then cover in section III many applications in the field of electron scattering on $^3$He ($^3$H). 
They are all based on the most simple nonrelativistic single nucleon current 
operator. 
It is only in section IV that we take into account some mesonic
 exchange currents via Siegert's theorem\cite{REF3}. 
As we shall see their effects on the tensor polarization in a pd capture process are very strong.
We conclude in section V.

\section{The Exact Treatment of FSI}

Once the photon is absorbed by the hadronic 3N system the three nucleons are no longer bound and scatter among each other. 
This is exemplified for the process $^3$He(e,e'p)d and based on a single nucleon current operator in Fig 1.
\begin{figure}
%\\[3mm]
\centerline{\mbox{\epsfysize=50mm \epsffile{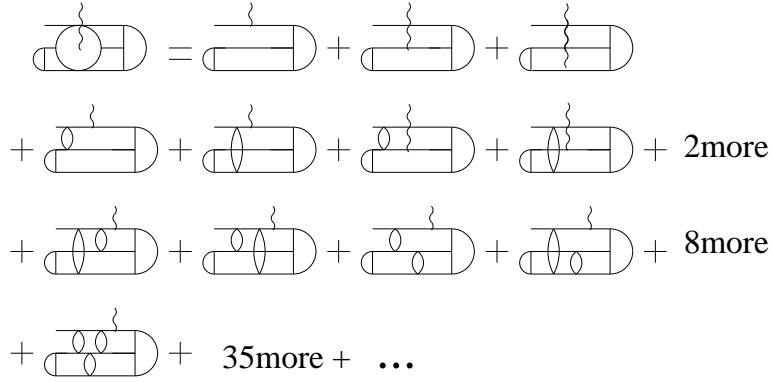}}}
%\\[3mm]
\caption{  The multiple rescattering series for the process $^3$He(e,e'p)d. The larger half moon stands for the $^3$He state, the wavy line for the photon, horizontal lines for freely propagating nucleons, the ovals for NN t-matrices and the small half moon for the final deuteron.
}
%\hspace{15mm}
%\\[4mm]
\end{figure}

The nuclear matrixelement
\begin{eqnarray}
N^{\mu} \equiv \langle \Psi ^{ (-)} _{ p d } \vert j ^{ \mu } 
(Q) \vert \Psi _{ ^3He} \rangle 
\end{eqnarray}
is expanded into a multiple scattering series in powers of the NN t-matrices acting among the three final nucleons. 
Obviously $\Psi ^{(-)} _{p d } $ is a 3N scattering state, 
$j^{\mu} (Q)  = \sum _{ i=1 } ^{3 } j^\mu _{(i)} (Q) $ the single nucleon current operator and $\Psi _{ ^3 He}$ the 3N bound state. 
In the first three diagrams the final nucleon and the deuteron do not interact.
We denote the first diagram as PWIA and the first three as PWIAS,
where ``S" stands for symmetrization of the final state.
Then come six processes where the final nucleon interacts once with the 
constituents of the deuteron.
The processes of second and third order are also indicated. 
This infinite series is often diverging and has therefore to be summed up into 
an integral equation.

We introduce the following notation : $\vert \phi \rangle $ 
is the final channel 
state composed of a deuteron and a free nucleon, $t_i \equiv t_{jk}$ 
the NN (off shell) t-matrix for the pair $jk$, $G_0$ the free 3N propagator, 
$j$ the current operator and $\Psi_b$ the 3N bound state. 
Then the rescattering terms in Fig. 1 can be written as 
\begin{eqnarray}
N_{rescatt} &\equiv& 
\langle \phi \vert t_3 G_0 j \vert \Psi _b \rangle +
\langle \phi \vert t_2 G_0 j \vert \Psi _b \rangle 
\cr
&+&
\langle \phi \vert ( t_2 G_0 t_3 G_0 +  t_3 G_0 t_2 G_0 +
t_3G_0 t_1 G_0 + t_2 G_0 t_1G_0 ) j \vert \Psi_b \rangle
+ \cdots
\end{eqnarray}
This can be formulated concisely with the help of an operator  
$P\equiv P_{12}P_{23} + P_{13}P_{23}  $, which is 
 the sum of a cyclical and an anticyclical  permutation:
\begin{eqnarray}
N_{rescatt}  &=& \langle \phi \vert P t_1 G_0 j \vert \Psi_b \rangle 
+ \langle \phi \vert P t_1 G_0 P t_1 G_0 j \vert \Psi_b \rangle + \cdots
\cr
&=& \langle \phi \vert P  \{ t_1 G_0 j \vert \Psi \rangle + ( t_1 G_0 P) t_1 G_0 j \vert 
\Psi_b \rangle 
\cr
&+& (t_1 G_0 P) (t_1 G_0P) t_1 G_0 j \vert \Psi \rangle + \cdots \}
\cr
& \equiv & \langle \phi \vert P \vert U \rangle   
\end{eqnarray}
It follows that the quantity $\vert U \rangle $ 
\begin{eqnarray}
\vert U \rangle \equiv t G_0 j \vert \Psi_b \rangle 
+ (tG_0P) tG_0 j \vert  \Psi_b \rangle + (tG_0P) (tG_0P) tG_0 j \vert \Psi_b 
\rangle + \cdots
\end{eqnarray}
obeys the integral equation 
\begin{eqnarray}
\vert U \rangle = t G_0 j \vert \Psi_b \rangle + tG_0P \vert U \rangle
\label{eq5}
\end{eqnarray}
which is of the Faddeev-type. The inhomogeneous term is driven by the NN t-matrix (we dropped the index 1), the current operator and the target bound state. 
This integral equation can be solved exactly for any type of realistic NN force\cite{REF2}. 
The nuclear matrixelement then takes the form ( up to a symmetrization factor of 3)
\begin{eqnarray}
N^\mu = \langle \phi \vert j \vert \Psi_b \rangle + 
\langle \phi \vert P \vert U \rangle
\end{eqnarray}

As a second example for treating FSI processes exactly we regard inclusive 
scattering, where response function to some operator $\hat O$ occur:
\begin{eqnarray}
R\equiv \sum _f \vert \langle f \vert \hat O \vert i \rangle \vert ^2 \delta
( \omega +E_i -E_f )
\end{eqnarray}
Here $E_i$ and $E_f$ are exact energy eigenvalues 
of the Hamiltonian $H$ to the 
initial and final states $\vert i \rangle $ and $\vert f \rangle $ and $\omega$ the energy carried by the photon. 
Apparently one can rewrite $R$ as
\begin{eqnarray}
R &=& - { 1 \over \pi } Im \sum _f \langle i  \vert \hat O ^\dagger \vert 
f \rangle { 1 \over { \omega +E_i -E_f + i\epsilon }} 
\langle f \vert \hat O \vert i \rangle 
\cr
&=&  - {1 \over \pi} Im \langle i \vert \hat O ^\dagger { 1 \over { \omega +E_i - H +i \epsilon }} \hat O \vert i \rangle
\end{eqnarray}
using the completeness relation. 
Standard steps \cite{REF2,REF4} applied to the 3N problem $( \omega + E_i -E_f + i \epsilon )^{-1} \hat O \vert i \rangle $ leads to 
\begin{eqnarray}
R= -{ 1 \over \pi } Im \langle i \vert \hat O ^\dagger ( 1 + P  ) G_0 \vert 
\tilde U \rangle
\end{eqnarray}
where $\vert \tilde U \rangle $ obeys the integral equation.
\begin{eqnarray}
\vert \tilde U \rangle = ( 1 + t G_0) \hat O (1) \vert i \rangle + t P G_0 \vert \tilde U \rangle 
\label{eq10}
\end{eqnarray}
We see the same integral kernel as in (\ref{eq5}). 
This result is formulated for the case $\hat O \equiv \sum_{i=1} ^3 \hat O (i)$.
Our algorithm to solve Eqs. (\ref{eq5}) and (\ref{eq10}) and the technicalities connected with the partial-wave decomposition are described in\cite{REF2}. The expressions for the observables in terms of the nuclear matrixelements can be found also in \cite{REF2}.

\section{Applications}

The results to be presented are based on a strictly nonrelativistic treatment and the simple single nucleon current operator:
\begin{eqnarray}
\langle {\vec p}~' \vert j ^0 \vert \vec p \rangle &=& 
F^n _1 ({\vec p}~' - \vec p ) \Pi ^n + 
F^p _1 ({\vec p}~' - \vec p ) \Pi ^p
  \cr
\langle{ \vec p}~' \vert \vec j \vert \vec p \rangle &=&
{ { {\vec p}~' + \vec p } \over {2 m} } \{ F^n _1 ({\vec p}~ ' - \vec p ) \Pi ^n 
+ F^p _1 ( {\vec p}~' - \vec p) \Pi ^p \}
\cr 
&+&
{ { i \sigma \times ( {\vec p}~ ' - \vec p )} \over { 2 m} } \{ 
(F^n _1({\vec p}~' -\vec p) + 2m F^n _2 ({\vec p}~ ' - \vec p)  )\Pi ^n 
\cr
&+&
(F^p _1({\vec p}~' - \vec p) + 2m F^p _2 ({\vec p}~' - \vec p) ) \Pi^p \}
\label{11}
\end{eqnarray}
Here $F_{1,2}$ are the electromagnetic nucleon form factors and $\Pi^{n,p} $ are neutron and proton projection operators.

\subsection{ Elastic scattering}

This is an old topic \cite{REF5} and our results based on the current given in Eq. (\ref{11}) can be found in\cite{KAMADA}. Throughout this overview we restrict ourselves to three momentum transfers of the photon $\vert \vec Q \vert \le $ 400MeV/c, where relativity is expected to play still a minor role. 
Within that limited interval the charge form factor is reasonably well described. For the magnetic form factor discrepancies develop towards the larger $\vert \vec Q \vert $ values, which are known to be cured by the action of mesonic exchange currents\cite{REF6}.

\subsection{ Inclusive Scattering }

From the experience in elastic scattering one has to expect that the longitudinal structure function $R_L$ should be fairly well predicted and deficiencies should show up in $R_T$ due to the missing MEC's. 
This is indeed the outcome \cite{REF2} as shown in Figs. 2 and 3  for $\vert \vec Q \vert $= 174, 250 and 300MeV/c. 

\begin{figure}
%\begin{figure}
%\vspace{5mm}
\mbox{\epsfysize=40mm \epsffile{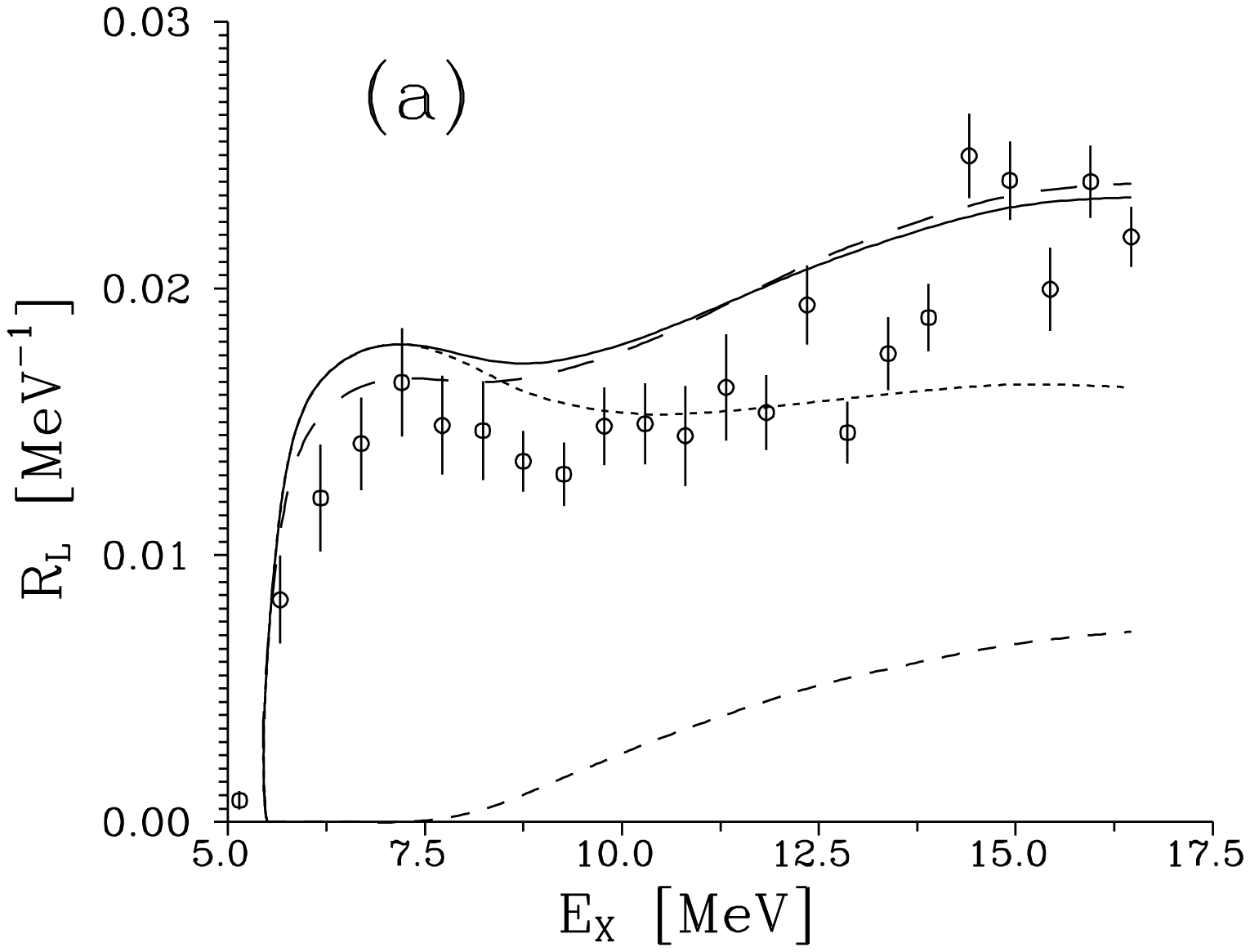}} \hspace{5mm}
\mbox{\epsfysize=40mm \epsffile{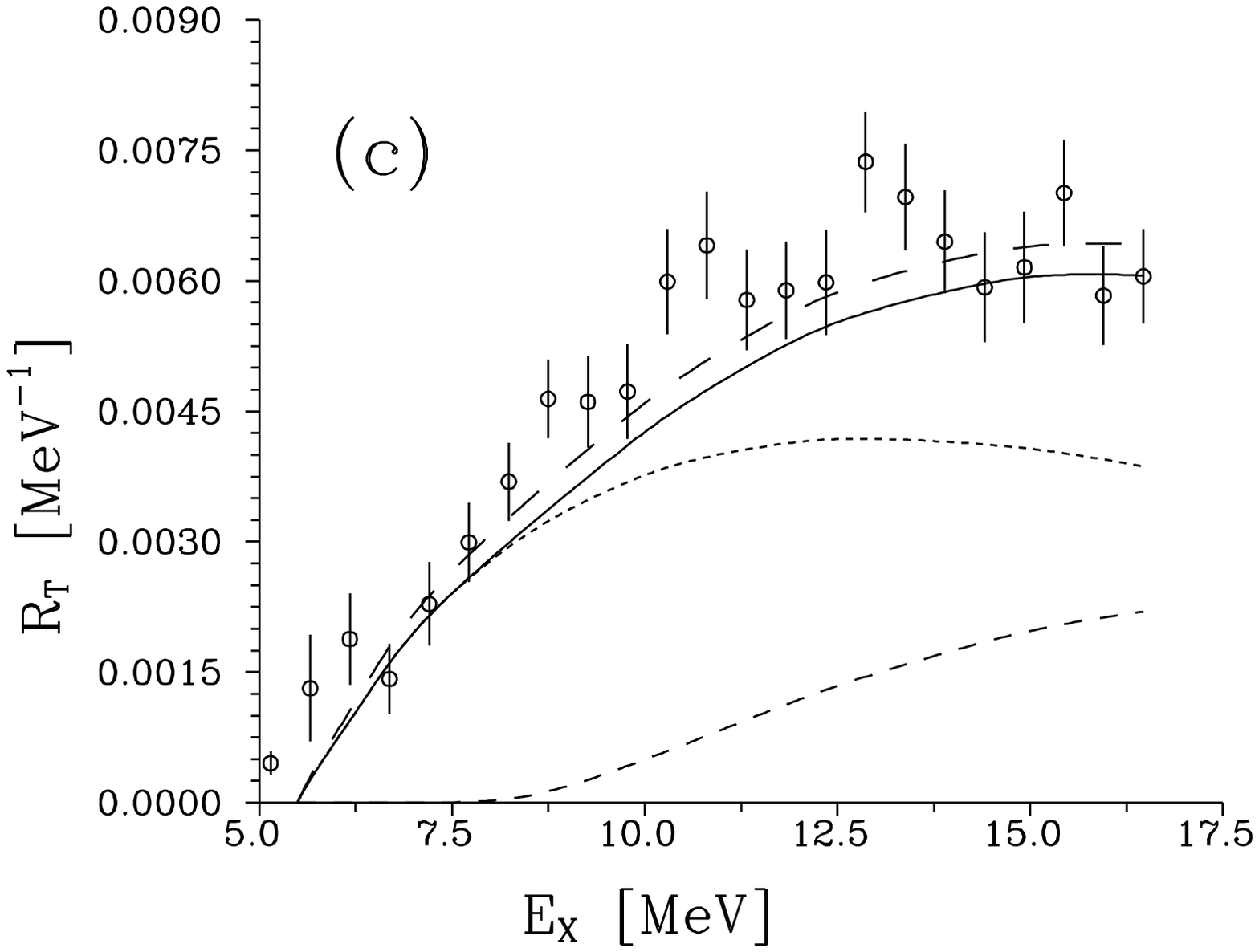}}
%\vspace{-60mm}
\mbox{\epsfysize=40mm \epsffile{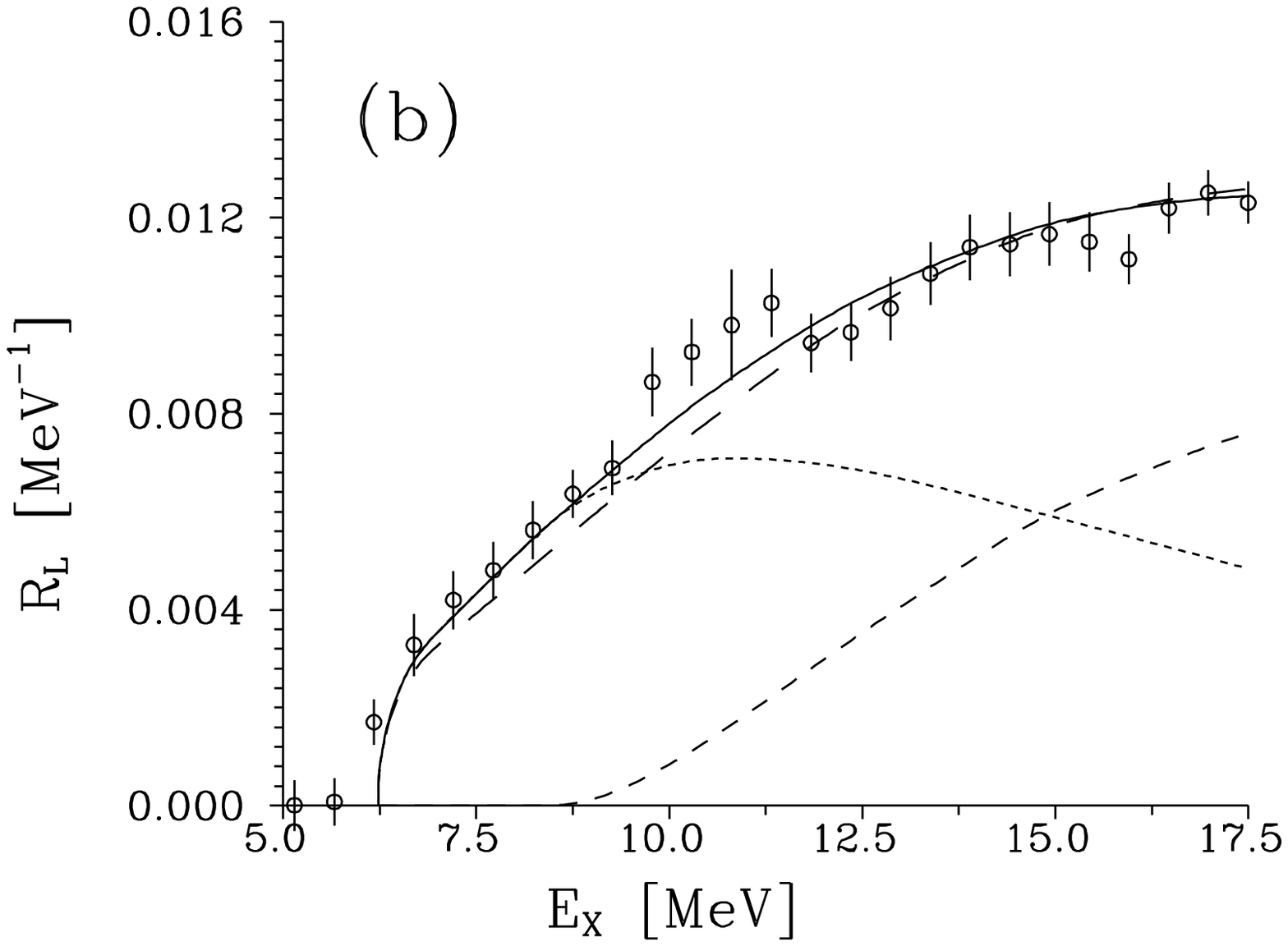}} \hspace{5mm}
\mbox{\epsfysize=40mm \epsffile{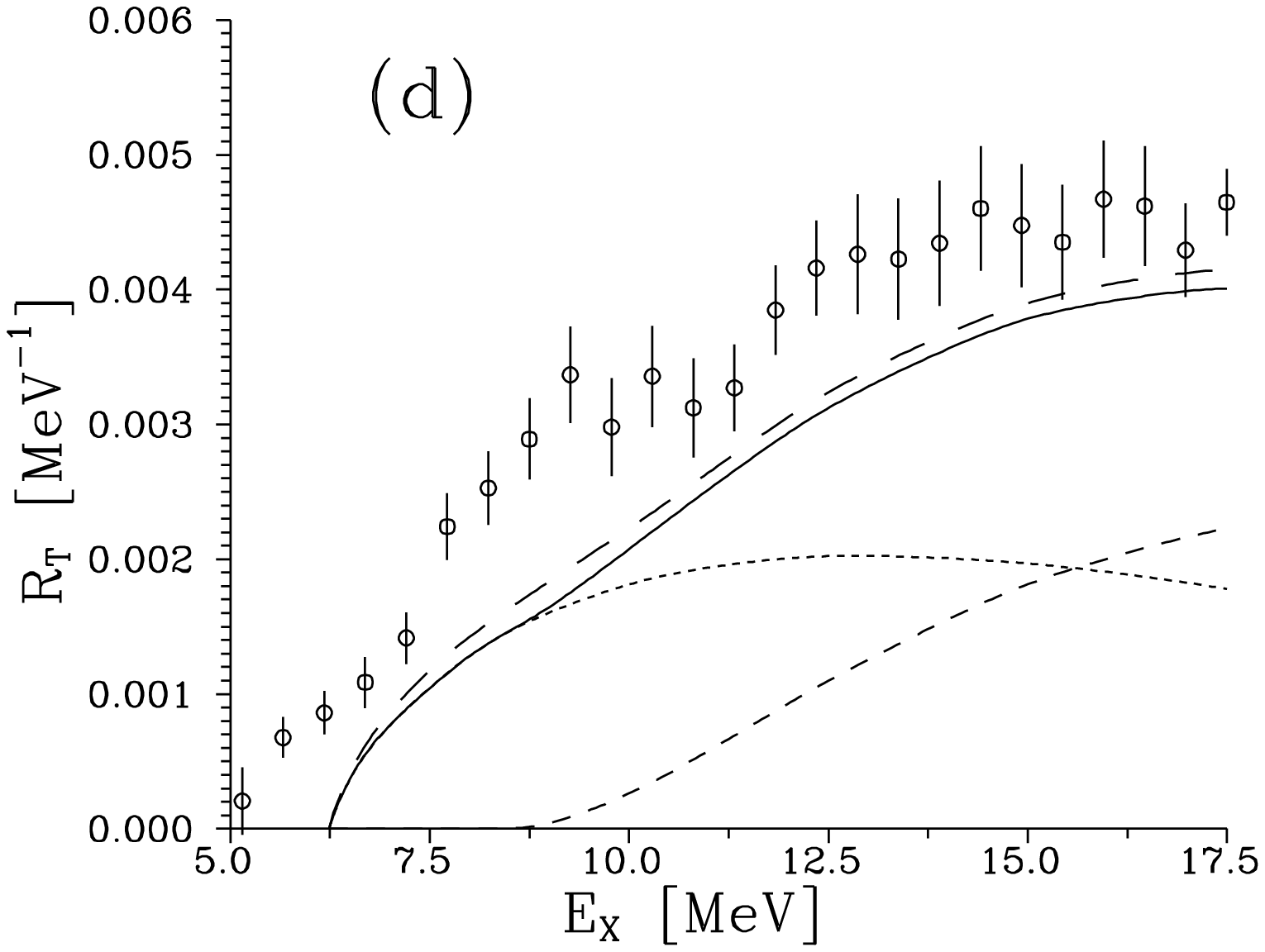}}
%
%\vspace{5mm}
\caption{ 
 (a) $^3$He longitudinal, (b) $^3$H longitudinal, (c) $^3$He transversal and 
(d) $^3$H transversal response functions at $Q$ = 174MeV/c. 
Comparison of data\protect\cite{REF7} to Bonn B\protect\cite{BONNB} (solid line) and MT I-III\protect\cite{MT} (long dashed line) calculations. 
Short dashed and medium long dashed lines are the separate contributions for the Nd and 3N breakups.  }
\end{figure}
\begin{figure}
\vspace{5mm}
\mbox{\epsfysize=40mm \epsffile{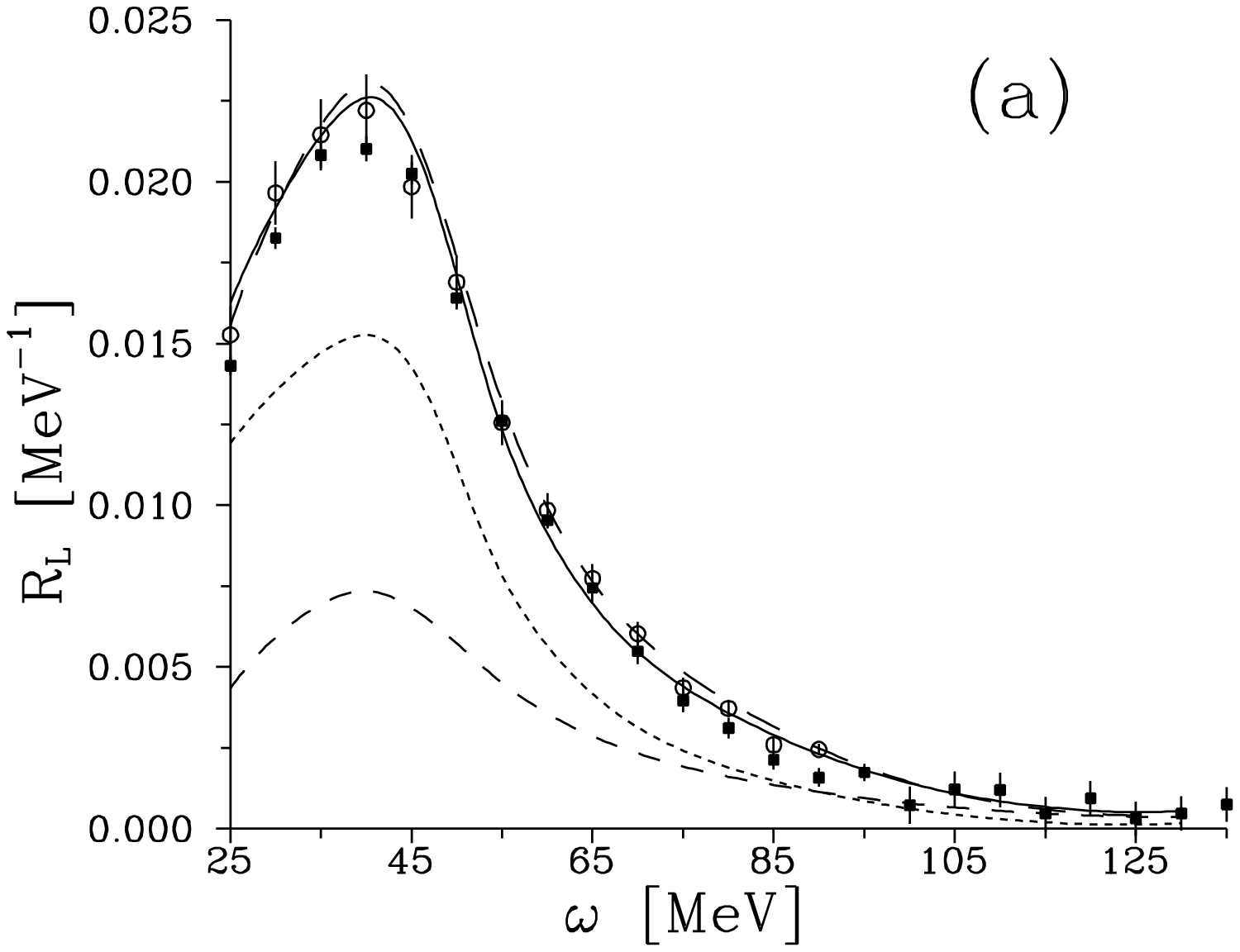}} \hspace{-5mm}
\mbox{\epsfysize=40mm \epsffile{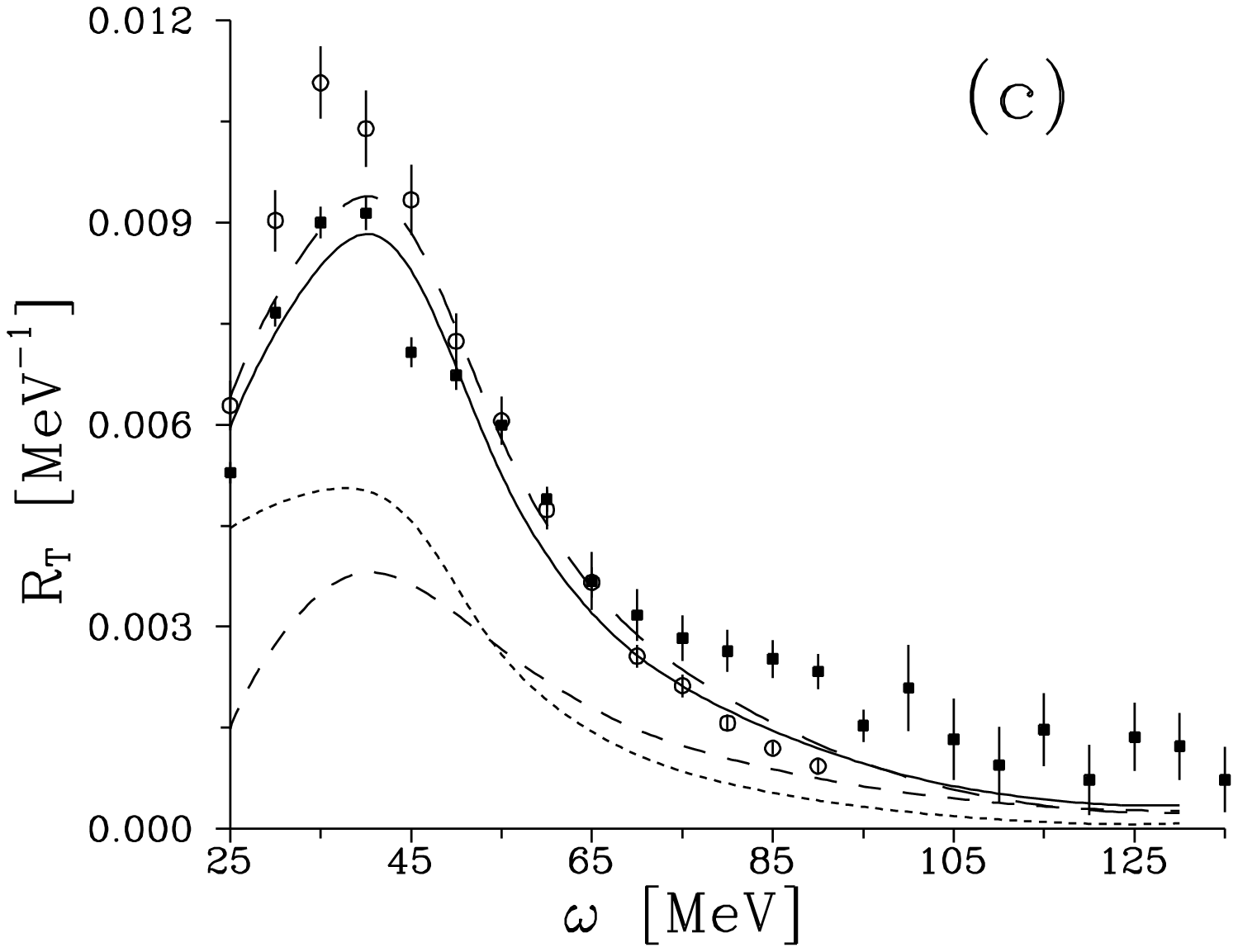}}
\\[3mm]
\mbox{\epsfysize=40mm \epsffile{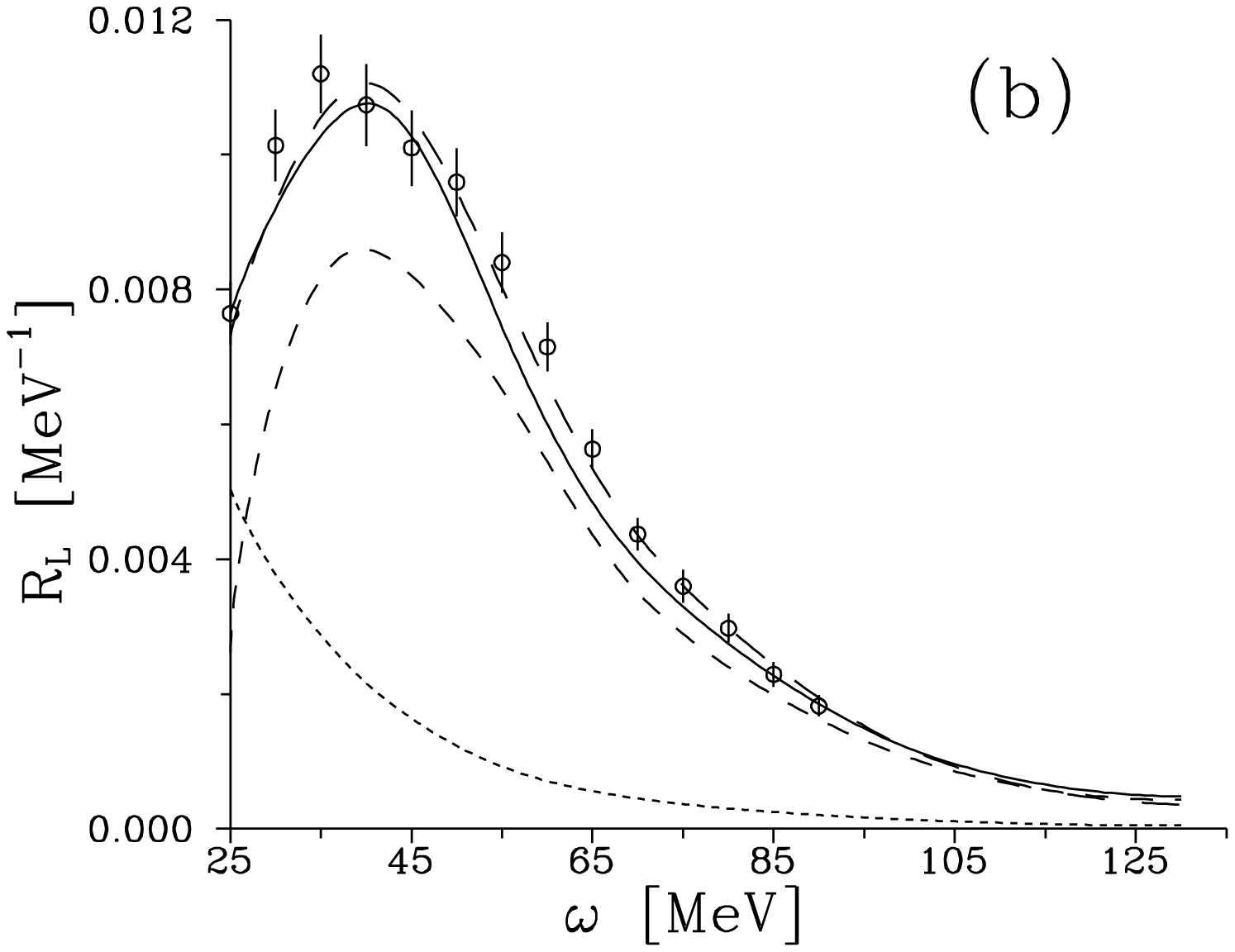}} \hspace{5mm}
\mbox{\epsfysize=40mm \epsffile{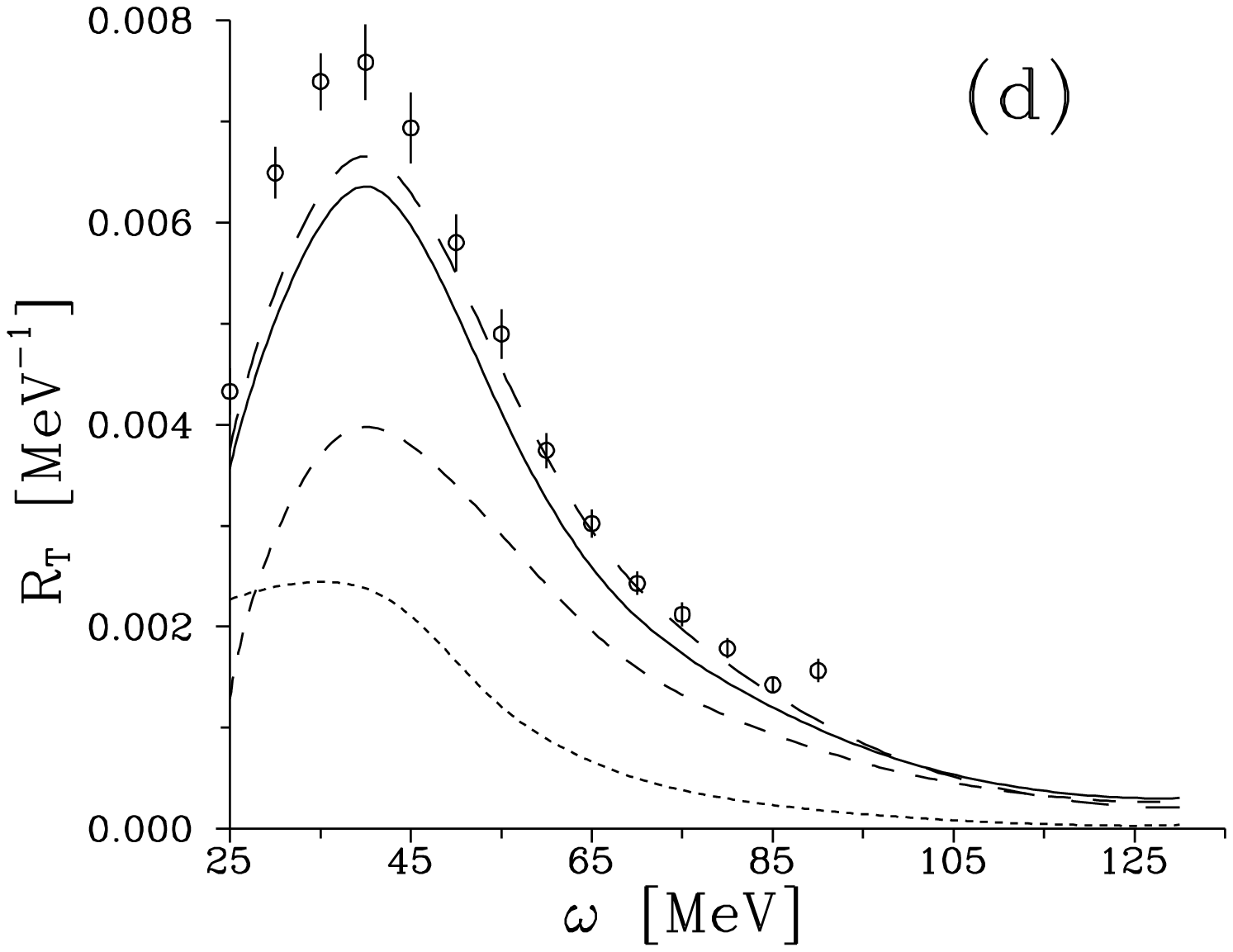}}
\vspace{10mm}
\caption{ 
 Same as in Fig. 2 for $Q$=250MeV/c. Data (circles) from \protect\cite{REF8} and (squares) from \protect\cite{REF9}.    }
\end{figure}

%\vfil
%\eject

%\begin{figure}
%\mbox{\epsfysize=40mm \epsffile{fig3a.ps}} \hspace{5mm}
%\mbox{\epsfysize=40mm \epsffile{fig3c.ps}}
%\\[3mm]
%\mbox{\epsfysize=40mm \epsffile{fig3b.ps}} \hspace{5mm}
%\mbox{\epsfysize=40mm \epsffile{fig3d.ps}}
%\\[10mm]
%\caption{ 
% Same as in Fig. 3 for $Q$=300MeV/c. }
%\hspace{15mm}
%%\\[2mm]
%\end{figure}               

\begin{figure}
\mbox{\epsfysize=40mm \epsffile{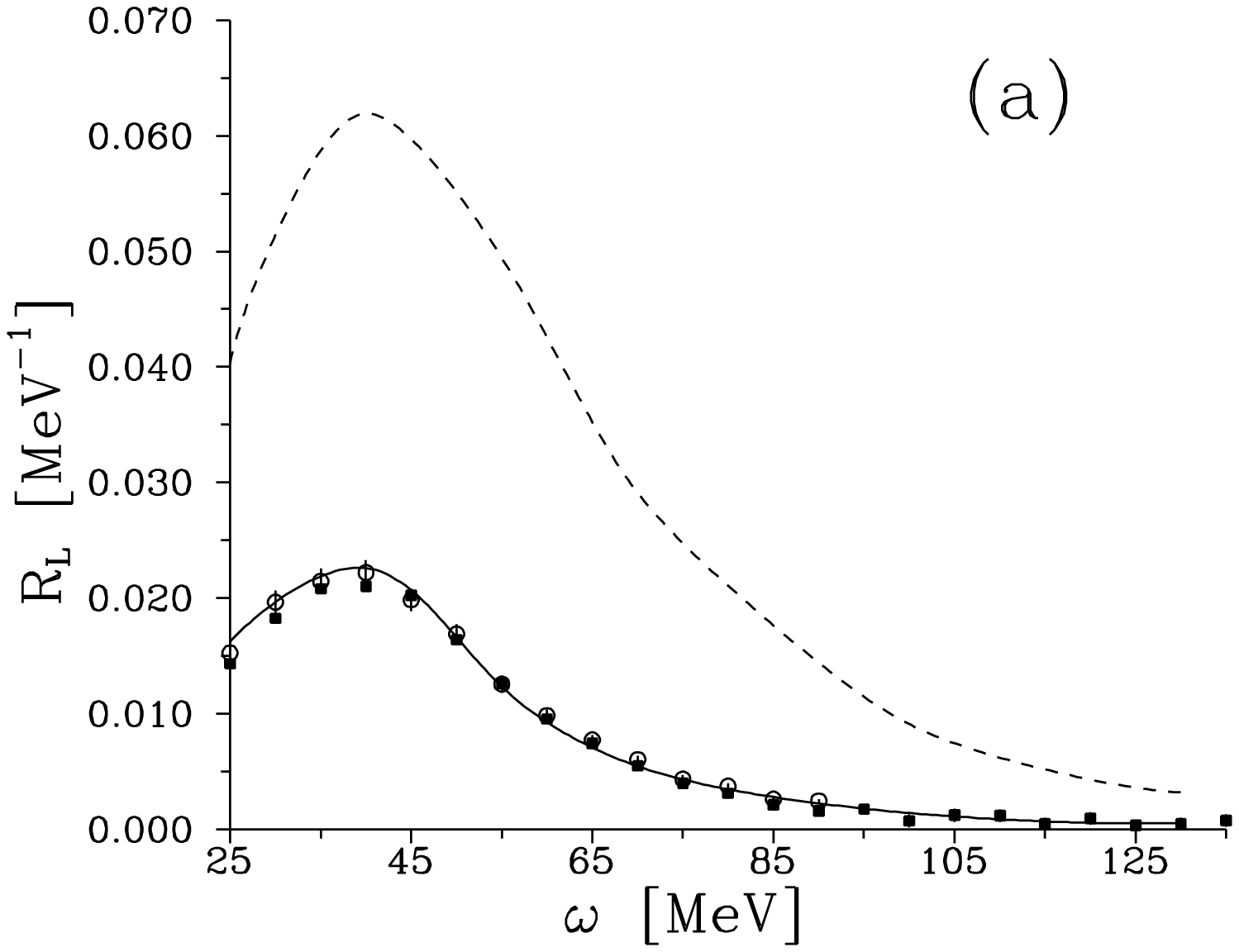}} \hspace{5mm}
\mbox{\epsfysize=40mm \epsffile{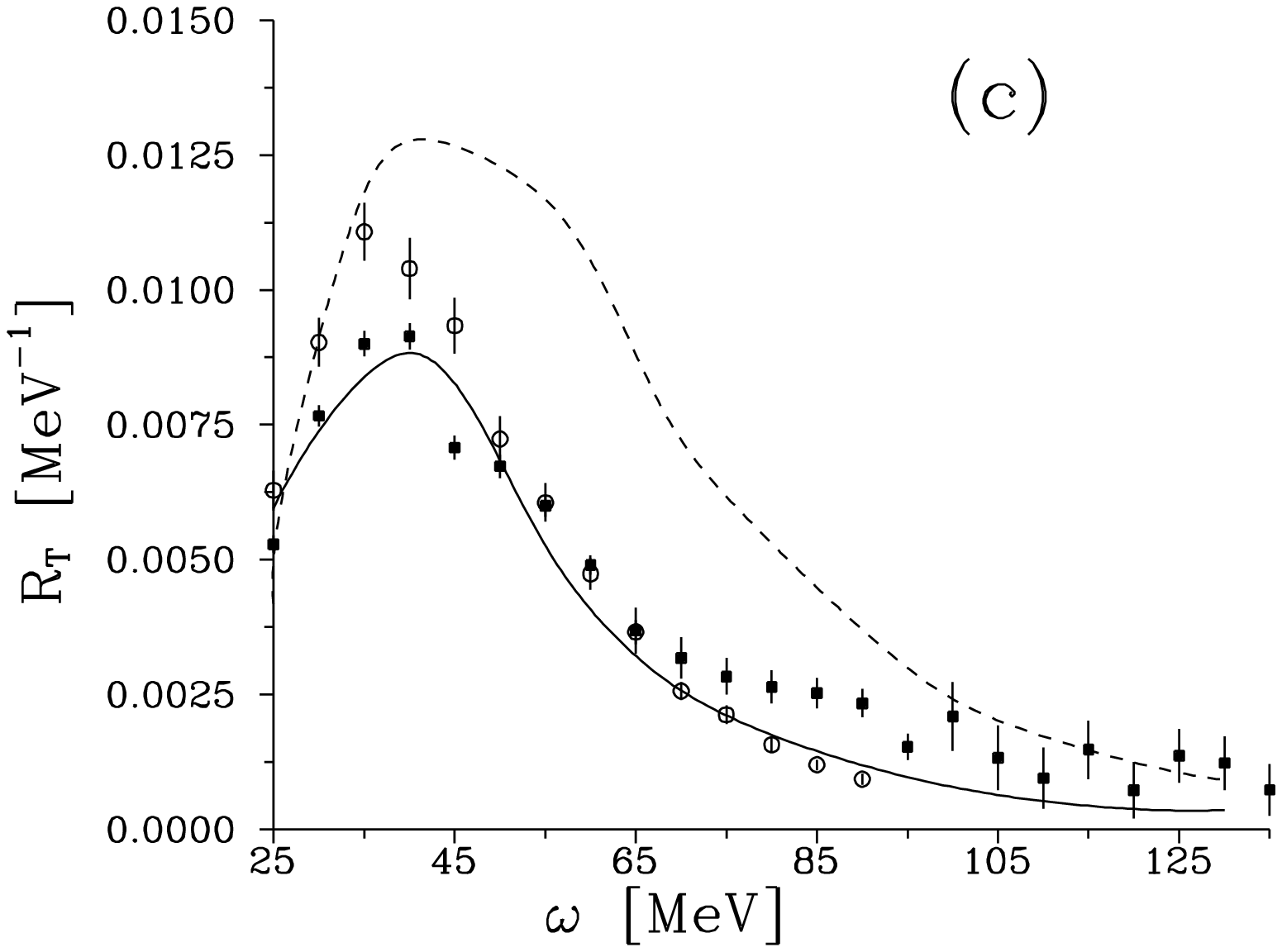}}
%\\[3mm]
\mbox{\epsfysize=40mm \epsffile{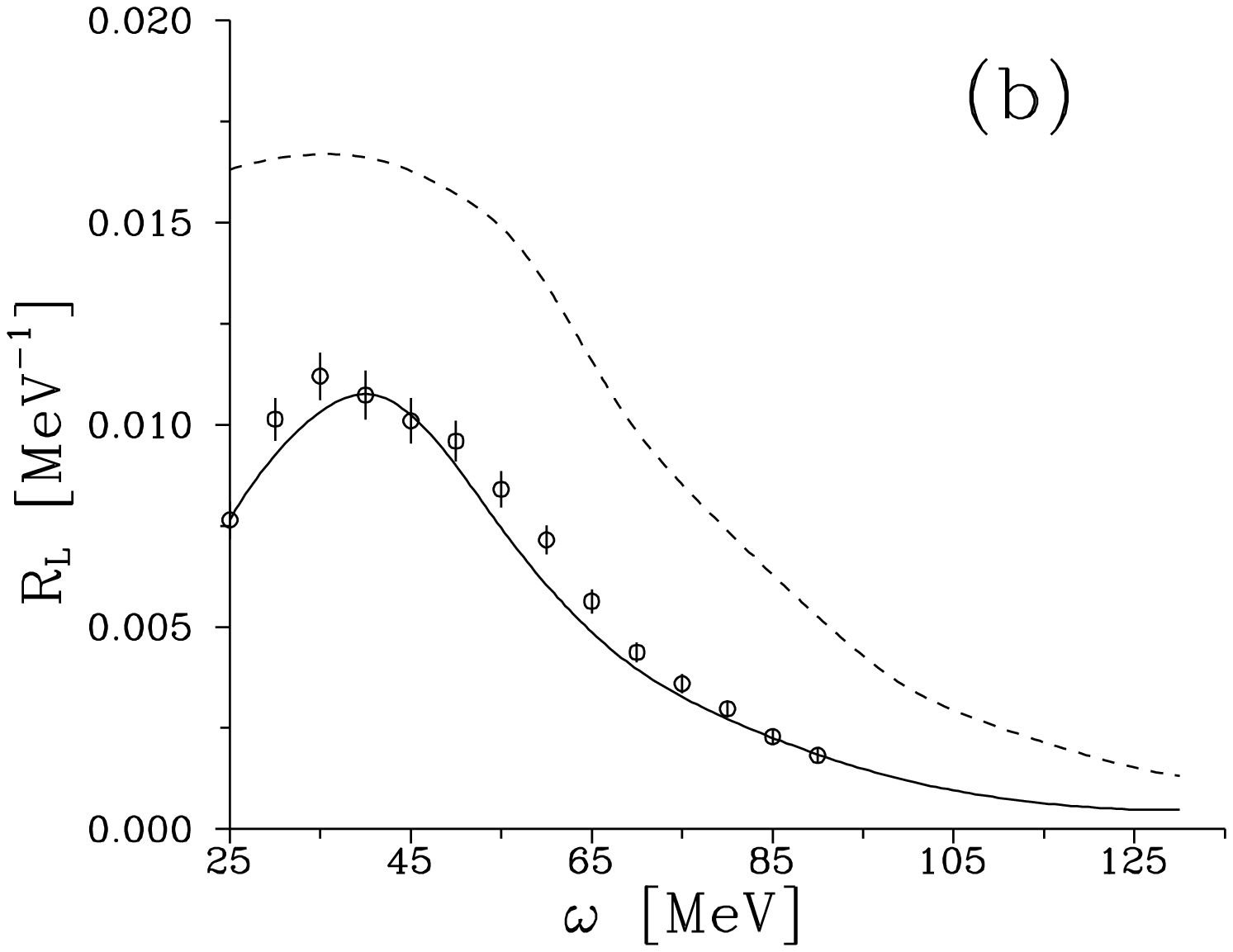}} \hspace{5mm}
\mbox{\epsfysize=40mm \epsffile{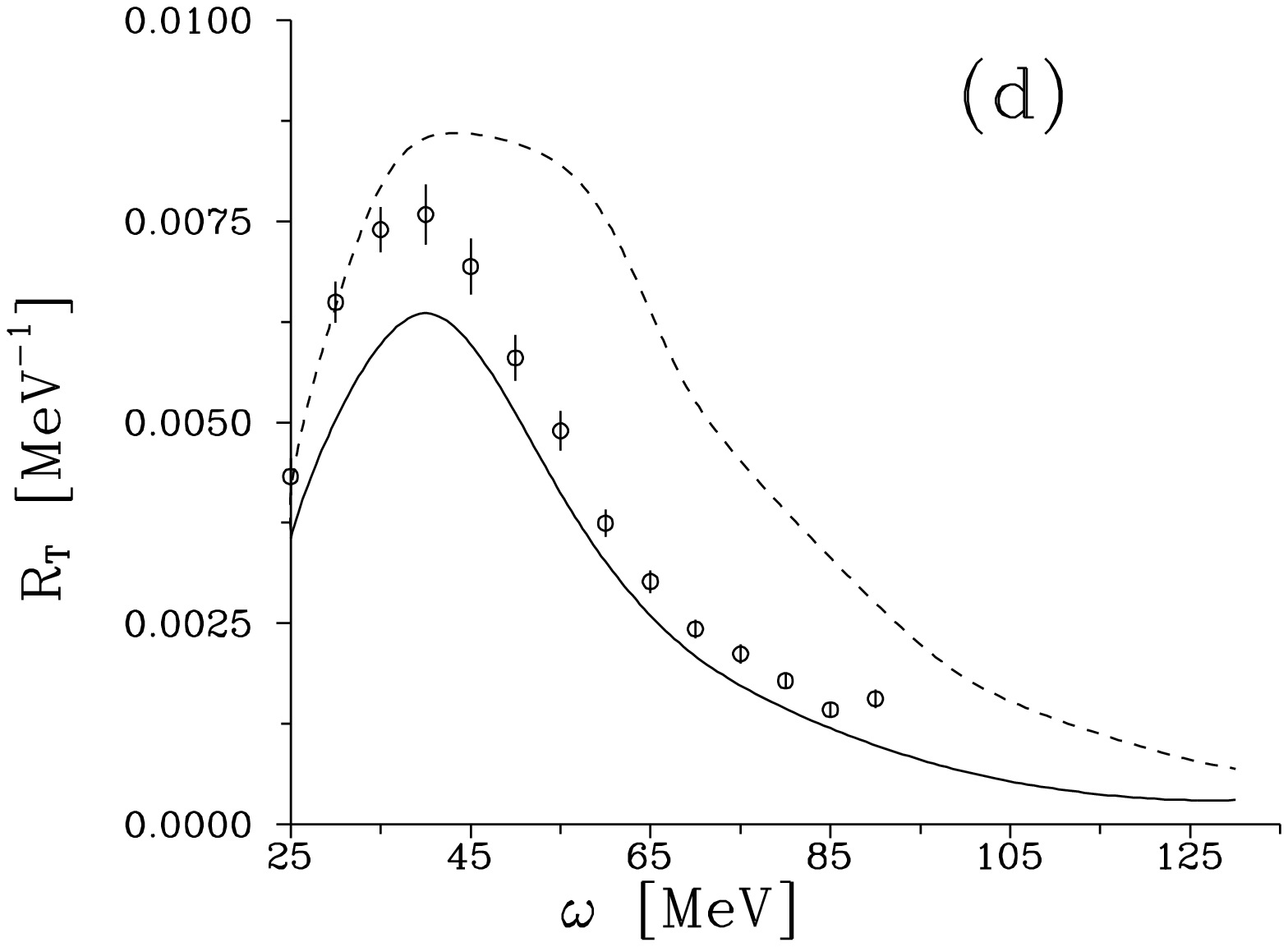}}
%\\[10mm]
\caption{ 
 (a) $^3$He longitudinal, (b) $ ^3$H longitudinal, (c) $^3$He transversal and 
(d) $ ^3$H transversal response functions at $Q$=250MeV/c. The PWIAS predictions (short dashed) are compared to the full Bonn B calculation and the data. }
\hspace{15mm}
%\\[2mm]
\end{figure}
%\vfil
%\eject

\begin{figure}
\mbox{\epsfysize=40mm \epsffile{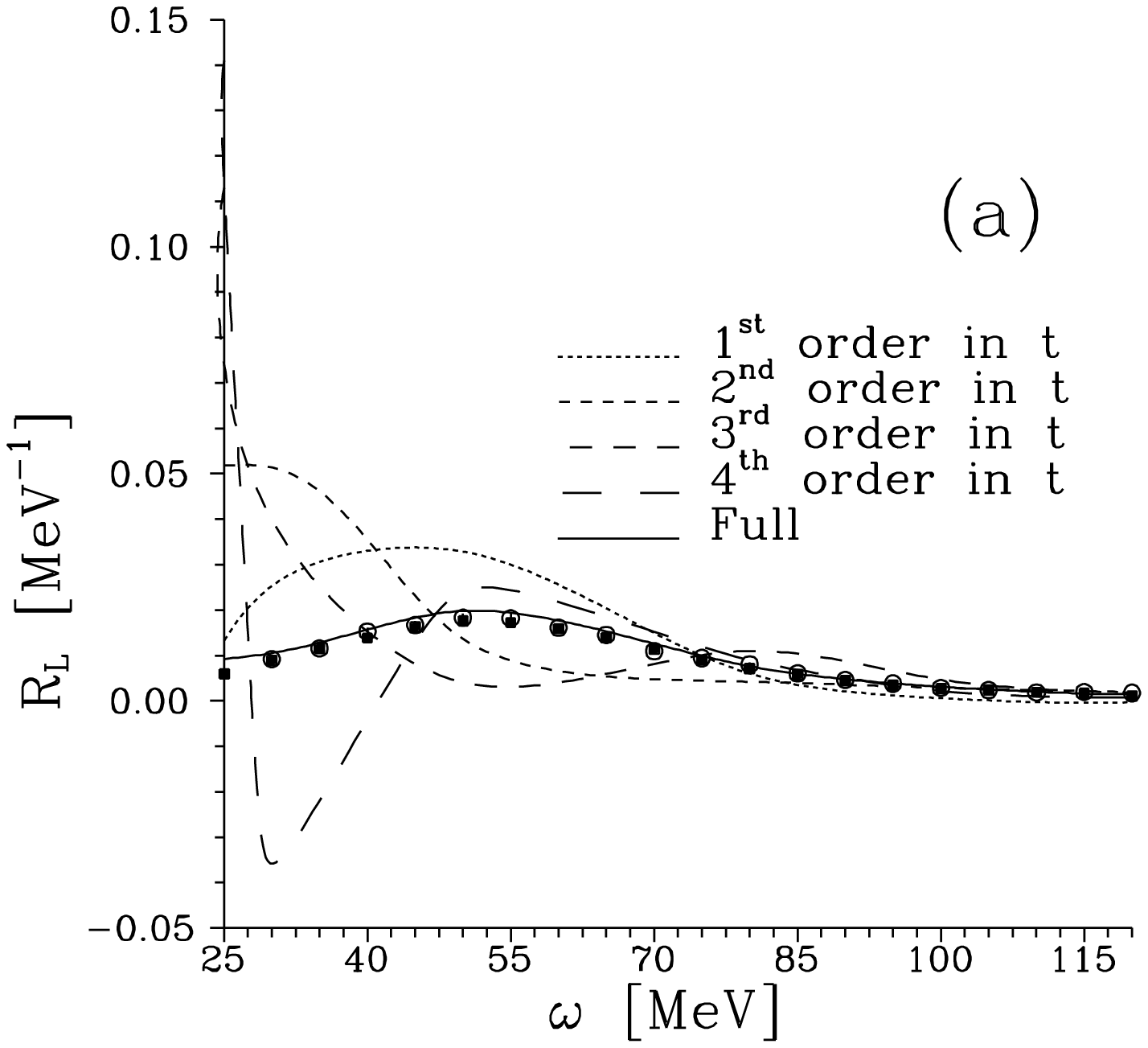}} \hspace{15mm}
\mbox{\epsfysize=40mm \epsffile{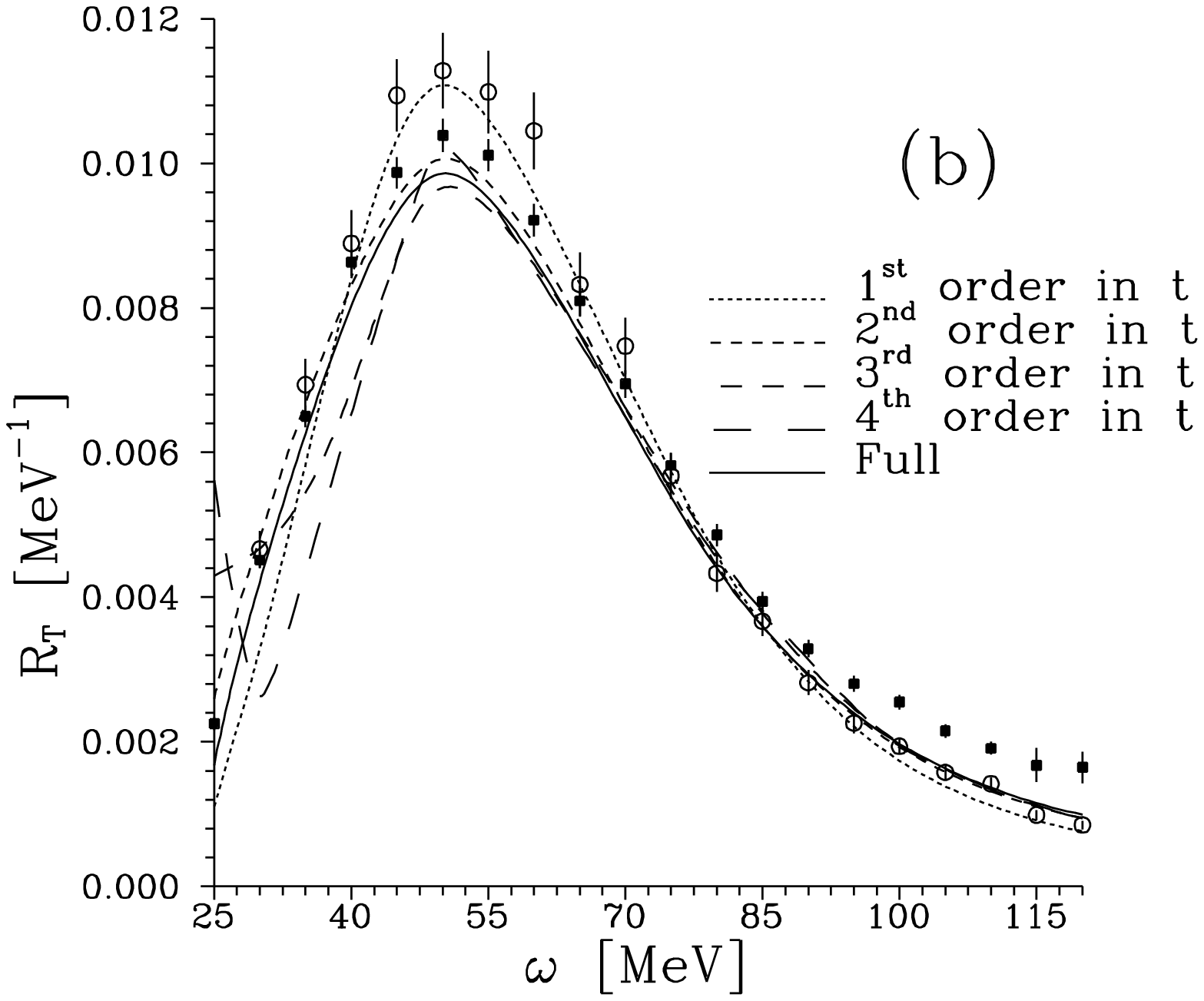}}
%\\[10mm]
\caption{  
The two responses for $^3$He, (a) longitudinal and (b) transversal at $Q$=300MeV/c. Shown are the PWIAS plus different orders of rescattering in the two-nucleon $t$-matrix. No Pad\'e has been used. $Full$ and data as in Fig. 3.
  }
\hspace{15mm}
\\[2mm]
\end{figure}

The treatment of FSI is crucial as is drastically demonstrated in Fig. 4.

In Fig. 5 the two structure functions are displayed based on rescattering included up to different orders in the NN t-matrix. 
It is interesting to see that it diverges for lower $\omega$'s for $R_L$ and that $R_T$ is very well described truncating the series (erroneously) at first order in $t$.

We would like to add some  results \cite{REF2} on the extraction of the pp correlation function from the Coulomb sum. 
We show in Fig. 6 the pp correlation functions in $^3$He for point and extended proton:
\begin{eqnarray}
C^{pp} _{point} (x) = { 1 \over 2 } \sum _M \langle \Psi_b M \vert 
\sum _{i \ne j } \Pi ^p (i) \Pi ^p (j)  \delta ( \vec x -( \vec r_i -\vec r_j)) \vert \Psi _b M \rangle 
\label{Cpoint}
\end{eqnarray}
\begin{eqnarray}
C^{pp} _{ext} (x) = { 1 \over 2 } \sum _M \int d \vec r \langle \Psi_b M \vert 
\sum _{i \ne j } \Pi ^p (i) \Pi ^p (j) F^p _1  ( \vec r + \vec x -\vec r_i ) 
F^p_1( \vec r - \vec r_j ) \vert \Psi _b M \rangle
\cr
\label{Cext}
\end{eqnarray}
with 
\begin{eqnarray}
F^p_1(\vec r) = { 1 \over{ (2 \pi) ^3 } } \int d \vec Q e^{ - i \vec Q  \vec r} F^p _1 (\vec Q)
\end{eqnarray}
We see a spread for different underlying NN potentials caused by their different short range repulsions. 
The nucleonic form factors fill the dip at short distances. 
These configuration space features are reflected again in a spread in the correlation function in momentum space (see Fig. 7), which is defined as
\begin{eqnarray}
C(\vert \vec Q \vert )= 
\int d \vec x e ^{ i \vec Q \vec x} C^{pp}_{ext} (\vec x)
\label{CQF}
\end{eqnarray}

\begin{figure}
\centerline{\mbox{\epsfysize=60mm \epsffile{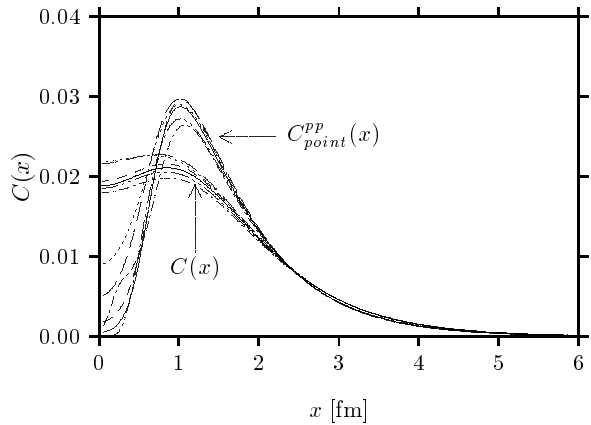}}} % \hspace{15mm}
%\vspace{100mm}
\caption{ 
 The two-nucleon correlation function $C(x)$ of Eq.(\protect\ref{Cext})
and the point proton-proton correlation function of Eq. (\protect\ref{Cpoint}) for various NN forces: AV18\protect\cite{AV18} (solid), Bonn B\protect\cite{BONNB} (dashed), Nijmegen93\protect\cite{NIJM93}(short dashed), Nijmegen I\protect\cite{NIJM93} (dotted), Paris\protect\cite{PARIS} (dash-dotted) and Ruhrpot\protect\cite{RUHRPOT} (dash-double-dotted).
  }
\end{figure}

It has been known for long time \cite{REF10} that the Coulomb sum is related to $C (\vert \vec Q \vert )$. If we define 
\begin{eqnarray}
S_L(\vert \vec Q \vert ) \equiv { 1 \over Z } \int _{\omega_{min}} ^\infty
d \omega R_L (\omega, \vert \vec Q \vert )
\end{eqnarray}
then under the assumption of a single nucleon density operator one finds 
(neglecting the neutron contribution)

\begin{eqnarray}
{ 1 \over Z } C ( \vert \vec Q \vert ) = S_L (\vert \vec Q \vert ) - 
\{ F_1 ^p (\vec Q ) \} ^2 + Z F^2 _{ch} (\vec Q)
\label{CQ}
\end{eqnarray}
where $F_{ch}$ is the charge form factor for elastic electron - $^3$He scattering. All four quantities entering Eq.(\ref{CQ}) are displayed in Fig. 8. 
One  sees the strong cancellation on the right hand side and the resulting $C ( \vert \vec Q \vert ) $ is roughly an order of magnitude smaller than $F^p _1$.
Apparently the control of $C$ will pose a challenge to experiment and theory.
Our simple theory for the left hand side of Eq. (\ref{CQF}) fails in comparison to the experimental values inserted into the right hand side, as seen in Fig. 9.

It has been shown in \cite{REF11} that relativistic corrections in the current operator and two-body current contributions remove most of the discrepancy.
These effects are large in relation to $C$ and no longer corrections. 
Clearly they call for a relativistic framework which might even change the interpretation and structure of the term $C$.

\begin{figure}
\vspace{5mm}
\centerline{\mbox{\epsfysize=60mm \epsffile{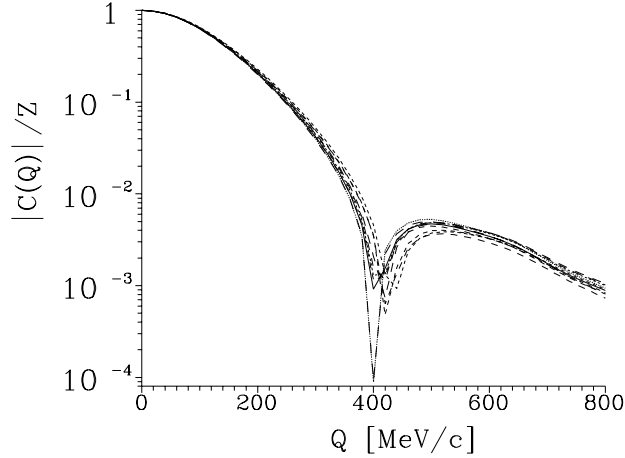}}}
\caption{
 The two-nucleon correlation function $\vert C(Q)\vert /Z$ of Eq. (\protect\ref{CQF}) for various NN forces. Description as in Fig. 6.
  }
\vspace{10mm}
\end{figure}

\begin{figure}
\vspace{10mm}
\centerline{\mbox{\epsfysize=60mm \epsffile{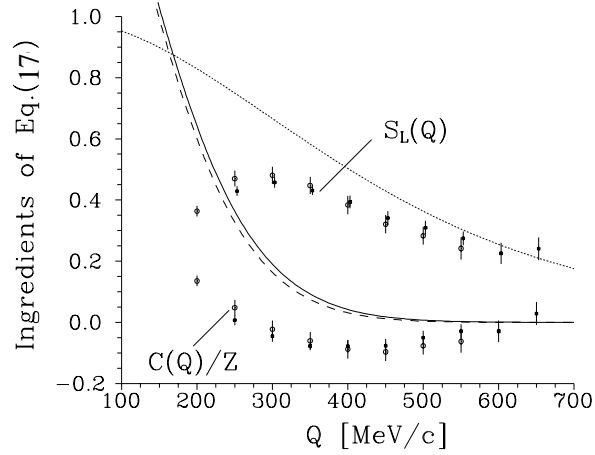}}} % \hspace{15mm}

\caption{ 
 The ingredients of Eq. (\ref{CQ}) for $^3$He: the experimental $S_L(Q)$
(open circles\protect\cite{REF8} and closed squares\protect\cite{REF9}) and [$ F^p _1 (Q)$]$^2$ from \protect\cite{GARI} (dotted) and $Z F_{ch}^2(Q)(1 - q _\mu ^2 / 4 M_{^3 He}^2) $
(experimental values \protect\cite{AMROUM})(solid curve), our theoretical values
 (dashed curve) and $C(Q)/Z$. }
\end{figure}
\begin{figure}
\centerline{\mbox{\epsfysize=60mm \epsffile{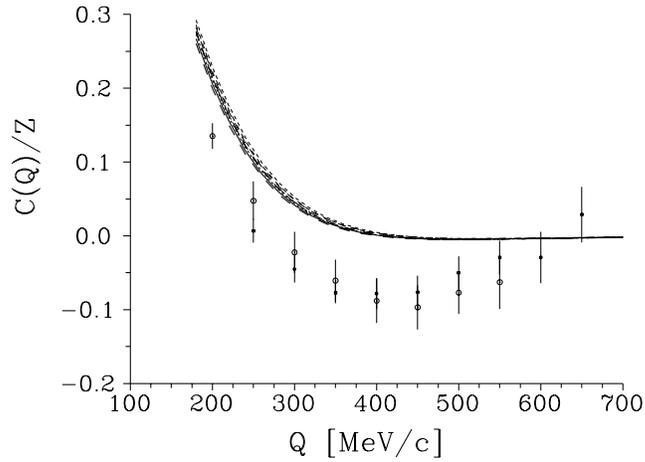}}}
%\\[10mm]
\hspace{25mm}
\caption{ 
Theoretical proton-proton correlation functions in $^3$He evaluated for various NN forces (as in Fig. 7) in comparison to the experimental correlation function $C(Q)/Z$ from Fig. 8. 
  }
\end{figure}

The access to $S_L$ requires extrapolation beyond the presently available data points for $R_L$. 
We show in Fig. 10  that theory might help. 
Putting 
\begin{eqnarray}
R_L( \omega ) = R_L (\omega _{max} ) ( { {\omega_{max}} \over {\omega}} )^\alpha
\label{TAIL}
\end{eqnarray}
we see that $\alpha$=4 is clearly favoured. 
This has been used in our analysis \cite{REF2} and we have found that 
the contributions of the extrapolated tails to the Coulomb sum range between 6 and 23 \%
in the case of the data set \cite{REF8} and even up to 40\% in case of the data\cite{REF9}.
Certainly for a future relativistic theory data less dependent on extrapolation assumptions would be desirable.

\begin{figure}
%\vspace{10mm}
\centerline{\mbox{\epsfysize=60mm \epsffile{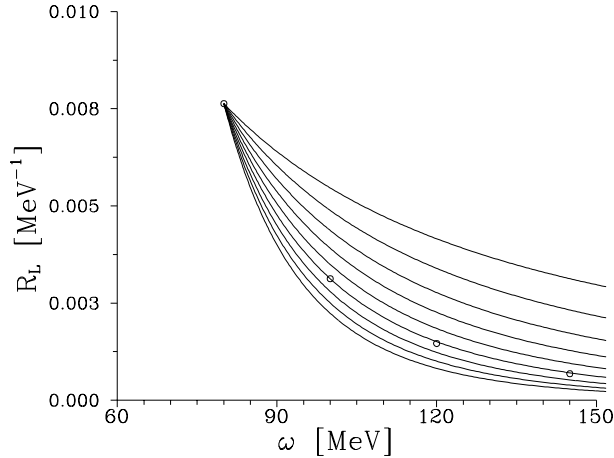}}}
%\\[10mm]

%\hspace{20mm}
\caption{ 
  The extrapolation according to Eq.(\protect\ref{TAIL}).
The exponent $\alpha$=4 is favored over $\alpha$=1.5, 2,2.5,3,3.5,4.5,5,5.5 and 6 read from top to bottom. The open circles are theoretical values at
$ Q$=300MeV/c.
  }
%\vspace{15mm}
\end{figure}

\subsection{ Exclusive scattering }

We regard the processes $^3$He(e,e'p)d and $^3$He(e,e'd)p, 
where data exist\cite{REF12} and new ones are coming up \cite{REF13}. 
To the best of our knowledge there are not yet data which cover the proton angular distribution over the whole proton knockout peak. 
One example of the rather limited  experimental knowledge is compared to theory 
in Fig. 11. 
The NN force (Bonn B) is kept different from zero in the states
$^1$S$_0$ and $^3$S$_1$-$^3$D$_1$, up to total angular momentum $j_{max}$=1 
and $j_{max}$=2, which clearly shows that the latter choice is sufficient.
Another example over a larger range of proton scattering angles is  shown in Fig. 12. 
In this example we see the importance of symmetrization in the final state (PWIAS) at large angles  and of FSI for all angles. 
Also it is important to use a realistic NN force and higher order rescattering processes have to be summed up correctly as is demonstrated in Figs. 13 and 14. 
The proton angular range around 200 $^\circ$ is where the deuteron knockout peak is located.
It can be seen as a peak  only by choosing the deuteron scattering angle. 
Unfortunately there are no data in the peak area. 
An example is displayed in Fig. 15.
Data\cite{REF12} right in the deuteron peak in the so called parallel kinematics 
$\vec {pd} \parallel \vec Q $ are shown in Fig. 16  in comparison to our theory. 
More data will come up \cite{REF13}. 
That knockout of a deuteron results from a complicated series of rescattering 
processes as dramatically illustrated in\cite{REF2}.

\begin{figure}
\vspace{10mm}
\centerline{\mbox{\epsfysize=60mm \epsffile{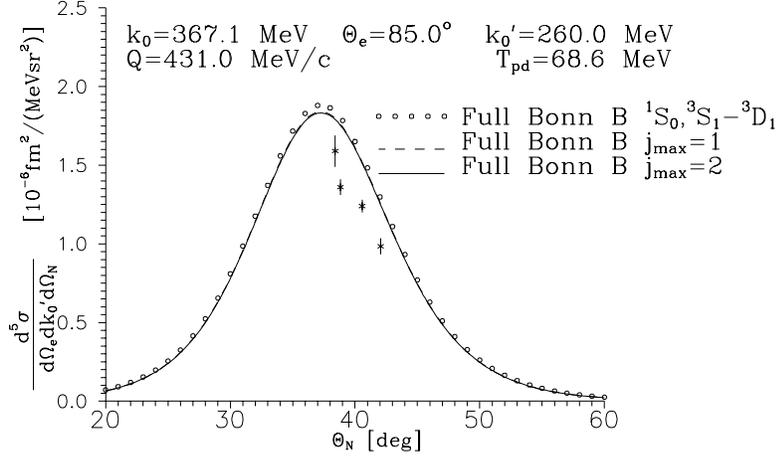}}} % \hspace{5mm} 
\caption{ 
 The quasifree proton knockout peak in $pd$ breakup. 
$\theta_N$ is the proton laboratory angle. The data are from 
\protect\cite{REF12}.
  }
\end{figure}
\begin{figure}
\centerline{\mbox{\epsfysize=60mm \epsffile{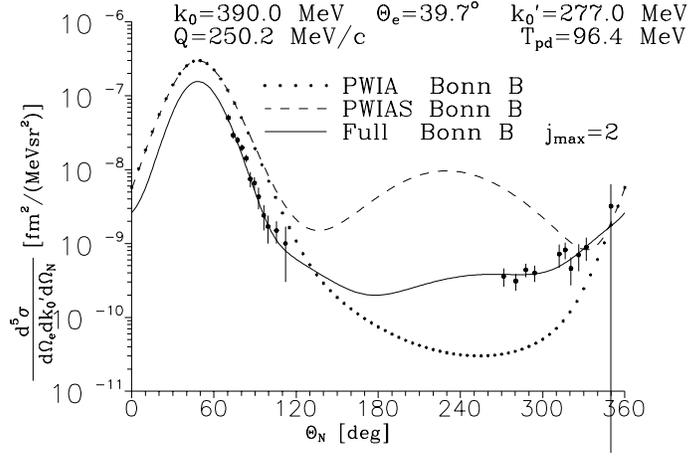}}}
\caption{ 
  The full angular proton distribution of $pd$ breakup. 
Comparison of PWIA, PWIAS, and the full treatment of rescattering. 
Data from \protect\cite{REF12}.
  }
\end{figure}
\begin{figure}
\vspace{10mm}
\centerline{\mbox{\epsfysize=60mm \epsffile{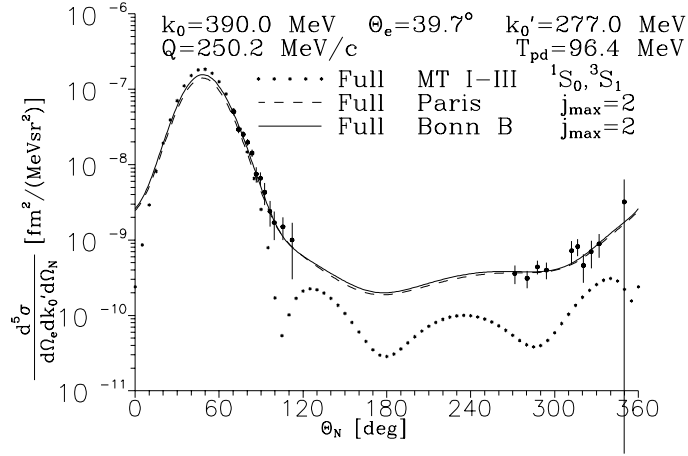}}} % \hspace{5mm}
\caption{ 
 Comparison of Paris and Bonn B predictions to the one based on the MT I-III\protect\cite{MT} $NN$ force model.
 }
\end{figure}

\begin{figure}
\vspace{10mm}
\centerline{\mbox{\epsfysize=60mm \epsffile{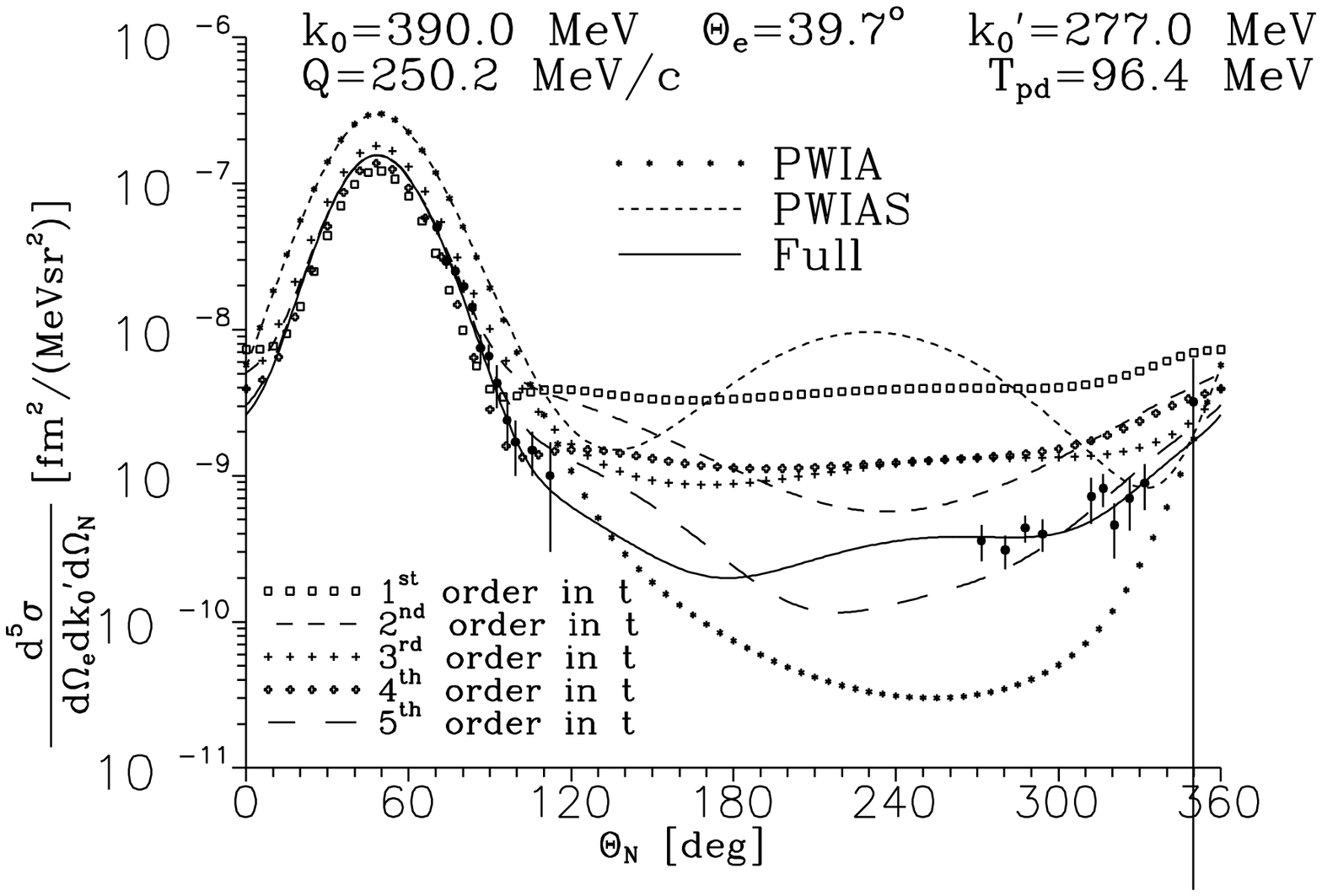}}}
\caption{ 
 Comparison of PWIA, PWIAS, and partial sums of increasing orders in $t$-matrix added to it.
 The convergence toward the full solution is very slow.
  }
\end{figure}

\subsection{ Inclusive Scattering with polarized electrons and $^3$He targets}

The process $\vec { ^3 He} ( \vec e , e' ) $ has been measured \cite{REF14,REF15,REF16} 
with the aim to extract the magnetic form factor of the neutron.
The basis for that possibility is the fact that the spin of a polarized $^3$He 
nucleus is carried to a very large extent by a polarized neutron \cite{REF17}.
The well known asymmetry expression 
\begin{eqnarray}
A= { { v_{T'} R_{T'} \cos {\theta ^* } + v_{TL'} R_{TL'} \sin{\theta  ^*}
\cos {\phi ^*} } \over { v_L R_L + v_T R_T }}
\end{eqnarray}
expressed in terms of the angle $\theta ^*$ between the direction of the $ ^3$He spin and the photon direction reduces to $A_{T'} $ for $\theta ^*$=0 and 
$A_{TL'}$ for $\theta ^*$=90 degrees. 
In PWIA and reducing the $^3$He state to the principle S-state (PS) one can show
\cite{REF2} that $A_{T'} \propto  ( G_M ^{(n)})^2 $ and $A_{T'L'} \propto
 G_E^{(n)} G_M ^{(n)}$. Of course this is no longer the case if
 FSI is included. 
Nevertheless one receives information on the magnetic form factor of the neutron\cite{REF2}. 
We compare in Figs. 17 and 18 various approximations to our best result and to the data. 
The most naive picture, PWIA and PS-state approximation of $^3$He fails totally for $A_{TL' } $ and is also a poor description for $A_{T'}$. 
Using the correct $^3$He state influences $ A_{TL'}$ very much but together 
with PWIA moves theory further away from the data.
Symmetrizing the final state according to PWIAS also influences $A_{TL'}$ strongly.
The partial inclusion of FSI among the two nucleons which are spectators to 
 the photon absorption(PWIA') causes 
a big effect, also the additional symmetrization(PWIAS'), but it is only after inclusion of the full FSI that one approaches the data both for $A_{T'}$ and $A_{TL'}$ 
fairly well.
More precise data for both asymmetries would be very desirable:
for $A_{T'}$ to better pin down $G_M ^{(n)}$ and for $A_{TL'}$ to probe 3N wave functions and the current operator.

\begin{figure}
\vspace{15mm}
\centerline{\mbox{\epsfysize=60mm \epsffile{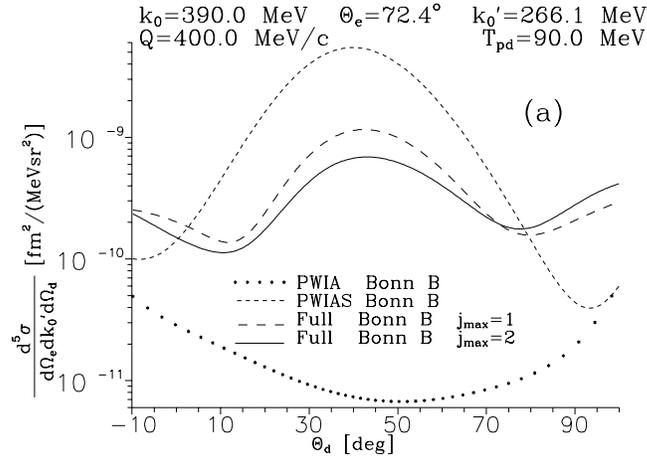}}} % \hspace{5mm}
\caption{ 
 A deuteron knockout peak.
Comparison of PWIA, PWIAS and two full calculations keeping the NN force up to
$j_{max}$=1 or 2.
  }
\end{figure}
\begin{figure}
\centerline{\mbox{\epsfysize=60mm \epsffile{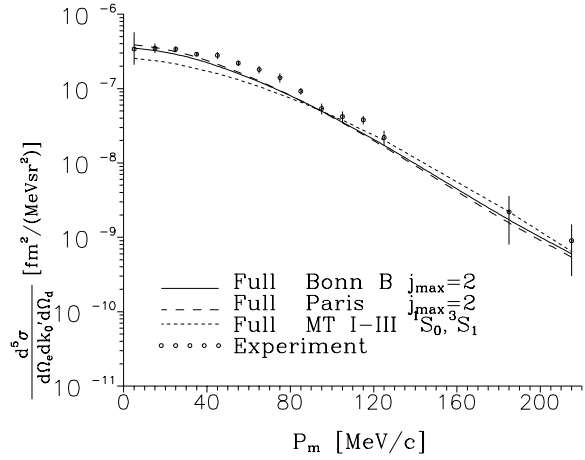}}}
\caption{ 
  Data\protect\cite{REF12}, in parallel kinematics against missing momentum $P_m$ are compared to the Paris\protect\cite{PARIS}, Bonn B\protect\cite{BONNB} and MT I-III\protect\cite{MT} predictions.
 }
\end{figure}
\begin{figure}
\vspace{15mm}
\input{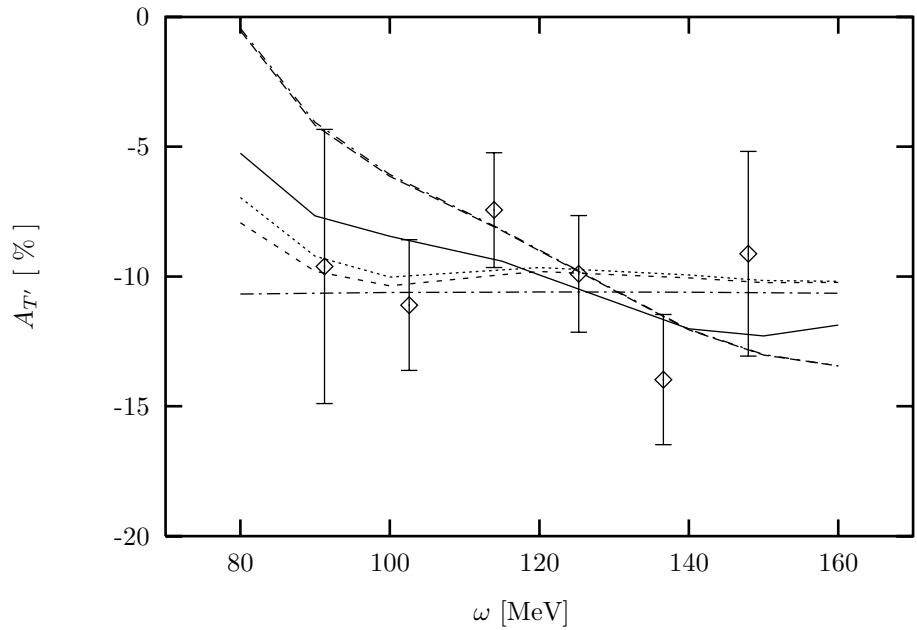}
\caption{ 
The transverse asymmetry $A_{T'}$ as a function of $\omega$. The data are from
Ref\protect\cite{REF14}. The six theoretical curves are
PWIA(PS)(dashed-dotted),
PWIA(dotted), PWIAS(short dashed), PWIA'(long dashed),
PWIAS'(dashed-dotted, declined curve) and FULL (solid). Note PWIA' and PWIAS'
overlap.
  }
\end{figure}
\begin{figure}
\vspace{15mm}
\input{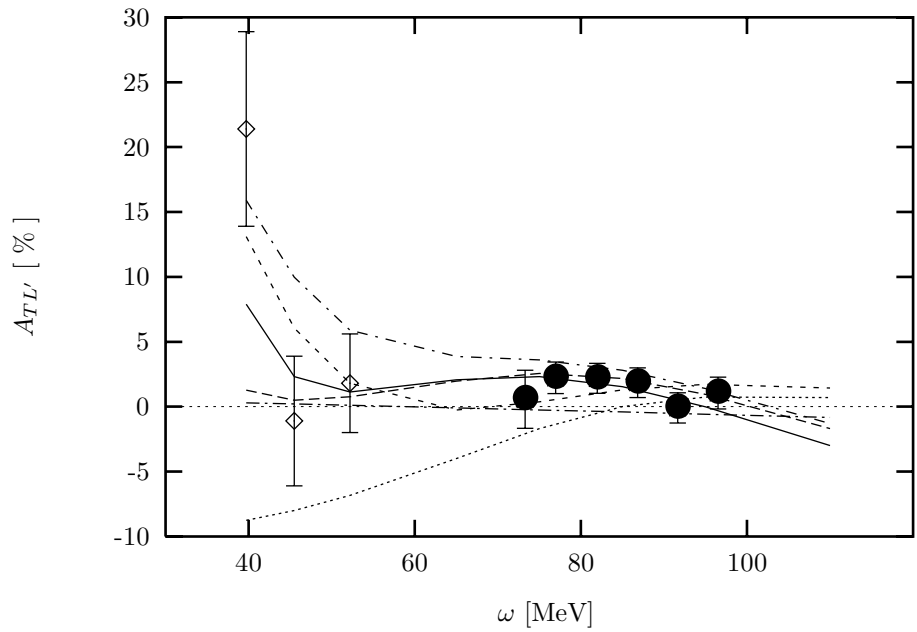}
\caption{ 
The transverse-longitudinal asymmetry $A_{TL'}$ as a function of $\omega$. The
data ($\diamond$) are from Ref\protect\cite{REF15} and the data
 ({\Huge $\bullet$}) from
Ref\protect\cite{REF16}. Curves as in Fig. 20.
The PWIAS'-curve rises to the data point at $\omega$=40MeV.
  }
\end{figure}

\subsection{ pd capture }

This example shows dramatically the improvement of a theoretical description of the data if the consistency between currents and Hamiltonian is taken into account according to Siegert's idea. 
We show in Fig. 19  the tensor analyzing power $A_{yy}$ for the process $\vec d + p \to \  ^3 He + \gamma $ at  $E_d$=45MeV\cite{REF18}.
The data are compared  to the single nucleon current prediction, which is far off and the dramatic improvement if a general current is included via the Siegert 
theorem in the electric multipoles. 
Fig. 20 demonstrates that the neglection of the initial state interaction would be a disaster. 
This is of course no surprise since Nd scattering at 45MeV requires the full multiple scattering series\cite{REF19}. 
Fig. 21 shows the contributions of various electric  multipoles  and 
points to the correct use of Siegert's replacement of parts of the current by 
the density operator. No long wave length approximation is required as will be shown in a forthcoming article. %\cite{REF20}.
Usually in the application of Siegert's theorem some terms are neglected,
illustrated here by the curve "$E_1$ one term ", whereas "$E_1, both~terms$" 
keeps correctly all terms. 
This observable $A_{yy}$ also reacts sensitively to switching off parts of the $^3$He wavefunction.
Dropping the D-state related amplitudes causes $A_{yy}$ to shift strongly away from  the data as shown in Fig 22. 
Finally we show in Fig. 23 the pd capture cross section itself  comparing in the single nucleon current approximation and using Siegerts theorem. 
As for $A_{yy}$ the latter formulation is by far superior also  for the cross section.

\begin{figure}
\vspace{15mm}
\centerline{\mbox{\epsfysize=60mm \epsffile{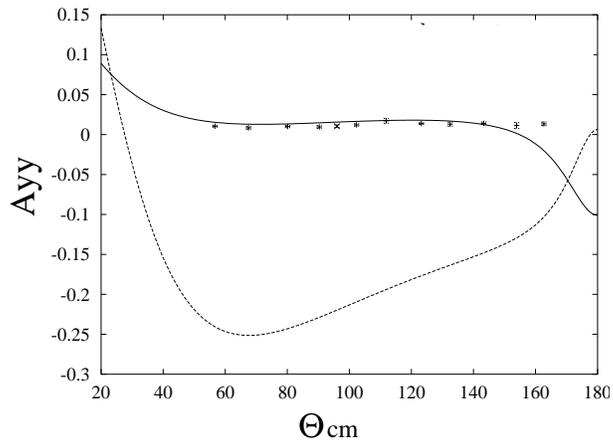}}} % \hspace{5mm}
\caption{ 
$A_{yy}$ with (solid curve) and without Siegert terms (dashed curve)
.
Data from\protect\cite{REF18} and the point at 96$^\circ$ from \protect\cite{JOURDAN}.
  }
\end{figure}

\begin{figure}
\centerline{\mbox{\epsfysize=60mm \epsffile{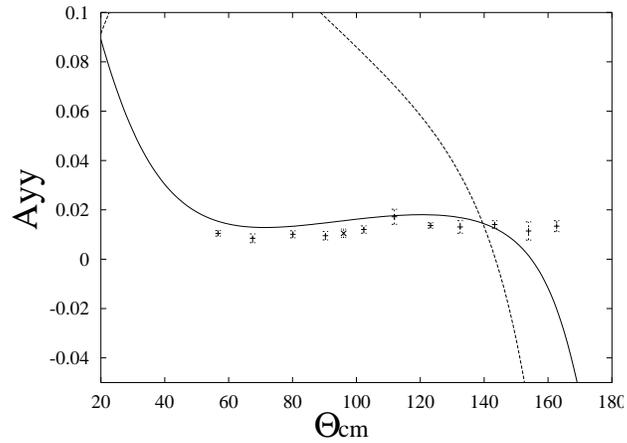}}}
\caption{ 
 The importance of the initial state interaction for $A_{yy}$ (solid curve) as compared to PWIAS (dashed curve). Data as in Fig. 19.
  }
\end{figure}

\begin{figure}
\centerline{\mbox{\epsfysize=60mm \epsffile{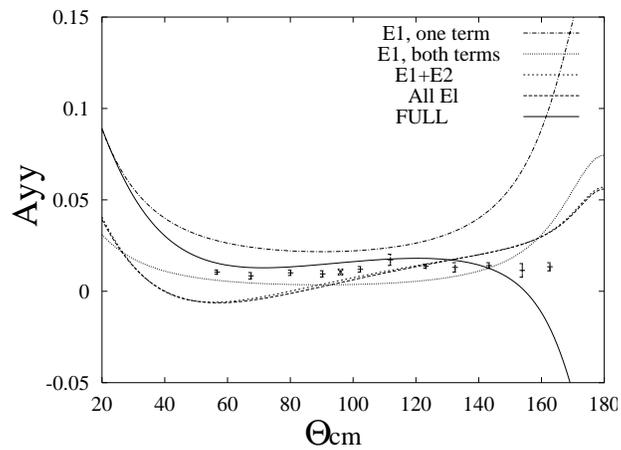}}} %\hspace{5mm}
\caption{ 
Various individual contributions to $A_{yy}$ in relation to our 
full result. Data as in Fig. 19. 
  }
\end{figure}

\begin{figure}
\centerline{\mbox{\epsfysize=60mm \epsffile{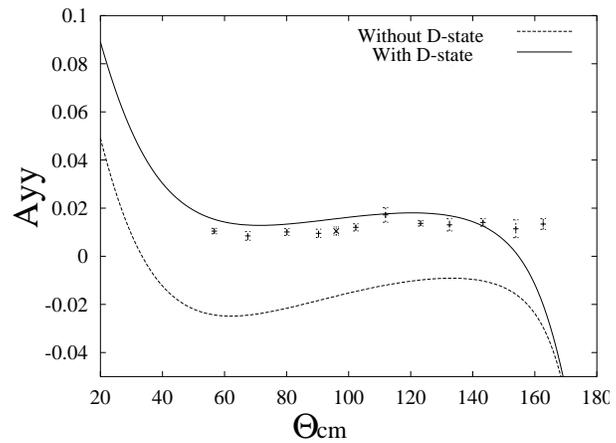}}}  % \hspace{15mm}
\caption{ 
Effect of D-wave admixture in the $^3$He bound state on $A_{yy}$. 
Data as in Fig. 19.}
\end{figure}
\begin{figure}
\mbox{\epsfysize=60mm \epsffile{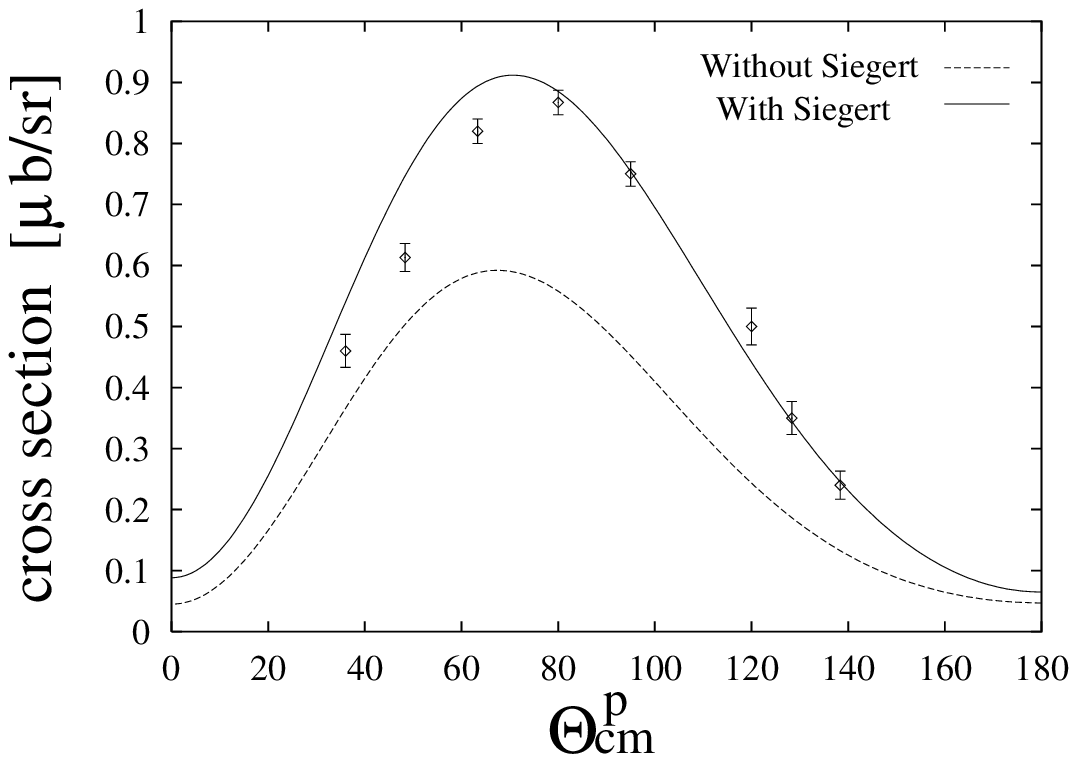}}
\caption{ 
Cross sections for the $pd$ capture reaction with and without Siegert terms. 
Data from \protect\cite{Angh}.
  }
\end{figure}

\section{ Conclusion }

We demonstrated that the correct treatment of the 3N continuum in various 
electromagnetic processes in the 3N system as well as the use of realistic NN 
forces is very important.
Though the restriction to a single nucleon current operator works often remarkably well for a convincing 
picture mesonic exchange currents have to be included and current conservation has to be fulfilled. 
For some of the observables which are accessible  by other techniques this has 
already been achieved with promising results \cite{REF6}. 
For exclusive processes much remains to be done. Using Siegert's idea in the pd capture process is a first step in that direction. 
We expect that 
the use of realistic forces together with a current operator consistent to them should be a sound basis to describe the data. 
First results shown in this small overview and restricted to small energy and momentum transfers of the photon are promising. 
At higher four momentum transfers relativity is required, which is a challenging task for theory.

%\bigskip
\section*{Acknowledgments}
This work was supported by
the Research Contract \# 41324878 (COSY-044) of the Forschungszentrum J\"ulich,
the Deutsche Forschungsgemeinschaft, 
the Polish Committee for Scientific Research under Grant No. 2P03B03914, and
the Science and Technology Cooperation Germany-Poland under Grant No.~XO81.91.
The numerical calculations have been performed on the Cray T90 of the
H\"ochstleistungsrechenzentrum in J\"ulich, Germany, and on the Convex 3820 of the Academic Computational Center (ACK) in Cracow, Poland. (KBN/SPP/UJ/019/1994).

\end{document}